\documentclass[aps,apl,twocolumn,groupedaddress,amsmath]{revtex4}
\usepackage[latin1]{inputenc}    
\usepackage{graphicx}   
\usepackage{xspace}

\begin{document}
%
\def\imat  {i}
\def\be{\begin{equation}}
\def\ee{\end{equation}}
\def\bea{\begin{eqnarray}}
\def\eea{\end{eqnarray}}
\def\la{\langle}
\def\ra{\rangle}
\def\l{\langle\!\langle}
\def\r{\rangle\!\rangle}

\def\Gin{\Gamma_\mathrm{in}}
\def\Gout{\Gamma_\mathrm{out}}
\def\Gs{\Gamma_\mathrm{S}}
\def\Gd{\Gamma_\mathrm{D}}
\def\Gc{\Gamma_\mathrm{C}}
\def\Grel{\Gamma_\mathrm{rel}}
\def\Gabs{\Gamma_\mathrm{abs}}
\def\Gem{\Gamma_\mathrm{em}}
\def\Gdet{\Gamma_\mathrm{det}}
\def\tdet{\tau_\mathrm{det}}
\def\Vtdeg{V_\mathrm{2DEG}}
\def\Iqpc{I_\mathrm{QPC}}
\def\Vqpc{V_\mathrm{QPC}}
\def\Gqpc{G_\mathrm{QPC}}
\def\Vsd{V_\mathrm{SD}}
\def\Vgl{V_\mathrm{G1}}
\def\Vgr{V_\mathrm{G2}}
\def\aGl{\alpha_\mathrm{G1}}
\def\aGr{\alpha_\mathrm{G2}}
\def\Idqd{I_\mathrm{DQD}}
\def\tc{t_\mathrm{C}}
\def\mul{\mu_\mathrm{1}}
\def\mur{\mu_\mathrm{2}}
\def\mus{\mu_\mathrm{S}}
\def\mud{\mu_\mathrm{D}}
\def\Ec{E_\mathrm{C}}
\def\Ecm{E_\mathrm{Cm}}
\def\Ecl{E_\mathrm{C1}}
\def\Ecr{E_\mathrm{C2}}
\def\Vqsd{V_\mathrm{QPC-SD}}
\def\Vdsd{V_\mathrm{DQD-SD}}
\def\Rs{R_\mathrm{S}}
\def\Rf{R_\mathrm{F}}
\def\Cc{C_\mathrm{C}}
\def\Cf{C_\mathrm{F}}
\def\Vout{V_\mathrm{out}}
\def\Vn{V_\mathrm{n}}
\def\Vn{I_\mathrm{n}}
\def\fBW{f_\mathrm{BW}}


\newcommand*{\eg}{e.\,g.\xspace}
\newcommand*{\ie}{i.\,e.\xspace}
\newcommand*{\cf}{c.\,f.\,,\xspace}
\newcommand*{\etc}{etc.\@\xspace}
\newcommand*{\etal}{\textsl{et al.\@\xspace}}

\newcommand*{\e}{\mathrm{e}}       
\newcommand*{\ci}{\mathrm{i}}
\newcommand*{\kb}{\ensuremath{\,k_\mathrm{B}}\xspace}
\newcommand*{\abs}[1]{\mid#1\mid}


\newcommand*{\um}{\ensuremath{\,\mu\mathrm{m}}\xspace}
\newcommand*{\nm}{\ensuremath{\,\mathrm{nm}}\xspace}
\newcommand*{\mm}{\ensuremath{\,\mathrm{mm}}\xspace}
\newcommand*{\m}{\ensuremath{\,\mathrm{m}}\xspace}
\newcommand*{\sqm}{\ensuremath{\,\mathrm{m}^2}\xspace}
\newcommand*{\sqmm}{\ensuremath{\,\mathrm{mm}^2}\xspace}
\newcommand*{\squm}{\ensuremath{\,\mu\mathrm{m}^2}\xspace}
\newcommand*{\psqm}{\ensuremath{\,\mathrm{m}^{-2}}\xspace}
\newcommand*{\psqmV}{\ensuremath{\,\mathrm{m}^{-2}\mathrm{V}^{-1}}\xspace}
\newcommand*{\cm}{\ensuremath{\,\mathrm{cm}}\xspace}

\newcommand*{\nF}{\ensuremath{\,\mathrm{nF}}\xspace}
\newcommand*{\pF}{\ensuremath{\,\mathrm{pF}}\xspace}

\newcommand*{\emob}{\ensuremath{\,\mathrm{m}^2/\mathrm{V}\mathrm{s}}\xspace}
\newcommand*{\edos}{\ensuremath{\,\mu\mathrm{C}/\mathrm{cm}^2}\xspace}
\newcommand*{\mbar}{\ensuremath{\,\mathrm{mbar}}\xspace}

\newcommand*{\A}{\ensuremath{\,\mathrm{A}}\xspace}
\newcommand*{\nA}{\ensuremath{\,\mathrm{nA}}\xspace}
\newcommand*{\pA}{\ensuremath{\,\mathrm{pA}}\xspace}
\newcommand*{\fA}{\ensuremath{\,\mathrm{fA}}\xspace}
\newcommand*{\uA}{\ensuremath{\,\mu\mathrm{A}}\xspace}

\newcommand*{\Ohm}{\ensuremath{\,\Omega}\xspace}
\newcommand*{\kOhm}{\ensuremath{\,\mathrm{k}\Omega}\xspace}
\newcommand*{\MOhm}{\ensuremath{\,\mathrm{M}\Omega}\xspace}
\newcommand*{\GOhm}{\ensuremath{\,\mathrm{G}\Omega}\xspace}

\newcommand*{\Hz}{\ensuremath{\,\mathrm{Hz}}\xspace}
\newcommand*{\kHz}{\ensuremath{\,\mathrm{kHz}}\xspace}
\newcommand*{\MHz}{\ensuremath{\,\mathrm{MHz}}\xspace}
\newcommand*{\GHz}{\ensuremath{\,\mathrm{GHz}}\xspace}
\newcommand*{\THz}{\ensuremath{\,\mathrm{THz}}\xspace}

\newcommand*{\K}{\ensuremath{\,\mathrm{K}}\xspace}
\newcommand*{\mK}{\ensuremath{\,\mathrm{mK}}\xspace}

\newcommand*{\kV}{\ensuremath{\,\mathrm{kV}}\xspace}
\newcommand*{\V}{\ensuremath{\,\mathrm{V}}\xspace}
\newcommand*{\mV}{\ensuremath{\,\mathrm{mV}}\xspace}
\newcommand*{\uV}{\ensuremath{\,\mu\mathrm{V}}\xspace}
\newcommand*{\nV}{\ensuremath{\,\mathrm{nV}}\xspace}

\newcommand*{\eV}{\ensuremath{\,\mathrm{eV}}\xspace}
\newcommand*{\meV}{\ensuremath{\,\mathrm{meV}}\xspace}
\newcommand*{\ueV}{\ensuremath{\,\mu\mathrm{eV}}\xspace}

\newcommand*{\T}{\ensuremath{\,\mathrm{T}}\xspace}
\newcommand*{\mT}{\ensuremath{\,\mathrm{mT}}\xspace}
\newcommand*{\uT}{\ensuremath{\,\mu\mathrm{T}}\xspace}

\newcommand*{\ms}{\ensuremath{\,\mathrm{ms}}\xspace}
\newcommand*{\s}{\ensuremath{\,\mathrm{s}}\xspace}
\newcommand*{\us}{\ensuremath{\,\mathrm{\mu s}}\xspace}
\newcommand*{\rpm}{\ensuremath{\,\mathrm{rpm}}\xspace}
\newcommand*{\minute}{\ensuremath{\,\mathrm{min}}\xspace}
\newcommand*{\degree}{\ensuremath{\,^\circ\mathrm{C}}\xspace}

\newcommand*{\EqRef}[1]{Eq.~(\ref{#1})}
\newcommand*{\FigRef}[1]{Fig.~\ref{#1}}

 \title{Detecting THz current fluctuations in a quantum point contact\\ using a nanowire quantum dot}
 \author{S.~Gustavsson}
 \email{simongus@phys.ethz.ch}
 \author{I.~Shorubalko} 
 \author{R.~Leturcq}
 \author{T.~Ihn}
 \author{K.~Ensslin}
 \affiliation {Solid State Physics Laboratory, ETH Zurich, CH-8093 Zurich,
 Switzerland}
 \author{S.~Sch\"on}
 \affiliation {FIRST Laboratory, ETH Zurich, CH-8093 Zurich,
 Switzerland}

\date{\today}
\begin{abstract}
%
We use a nanowire quantum dot to probe high-frequency current
fluctuations in a nearby quantum point contact. The fluctuations
drive charge transitions in the quantum dot, which are measured in
real-time with single-electron detection techniques.
The quantum point contact (GaAs) and the quantum dot (InAs) are
fabricated in different material systems, which indicates that the
interactions are mediated by photons rather than phonons. The large
energy scales of the nanowire quantum dot allow radiation detection
in the long-wavelength infrared regime.
\end{abstract}

\maketitle



Charge detection with single-electron precision provides a
highly-sensitive method for probing properties of mesoscopic
structures.
If the detector bandwidth is larger than the timescale of the
tunneling electrons, single-electron transitions may be detected in
real-time. This allows a wealth of experiments to be performed, like
investigating single-spin dynamics \cite{elzermanNature:2004},
probing interactions between charge carriers in the system
\cite{gustavsson:2005} or measuring extremely small currents
\cite{bylander:2005, fujisawa:2006, gustavssonAPL:2008}.
The quantum point contact (QPC) is a convenient detector capable of
resolving single electrons \cite{field:1993}.
Recently, it has been shown that the QPC not only serves as a
measurement device but also induces back-action on the measured
system \cite{onacQD:2006,khrapai:2006,gustavssonPRL:2007}. The
concepts of detector and measured system can therefore be turned
around, allowing a mesoscopic device like a quantum dot (QD) to be
used to detect current fluctuations in the QPC at GHz frequencies
\cite{aguado:2000}.

In this work we investigate a system consisting of a QPC defined in
a GaAs two-dimensional electron gas (2DEG) coupled to an InAs
nanowire QD.
We first show how to optimize the charge sensitivity when using the
QPC as single-electron detector. Afterwards, the system is tuned to
a configuration where electron tunneling is blocked due to Coulomb
blockade.
With increased QPC voltage bias we detect charge transitions in the
QD driven by current fluctuations in the QPC. The fact that the QPC
and the QD are fabricated in different material systems makes it
unlikely that the interactions are mediated by phonons
\cite{khrapai:2006}. Instead, we attribute the charge transitions to
absorption of photons emitted from the QPC \cite{aguado:2000,
gustavssonPRL:2007,zakka:2007}.

\begin{figure}[b]
\centering
 \includegraphics[width=\columnwidth]{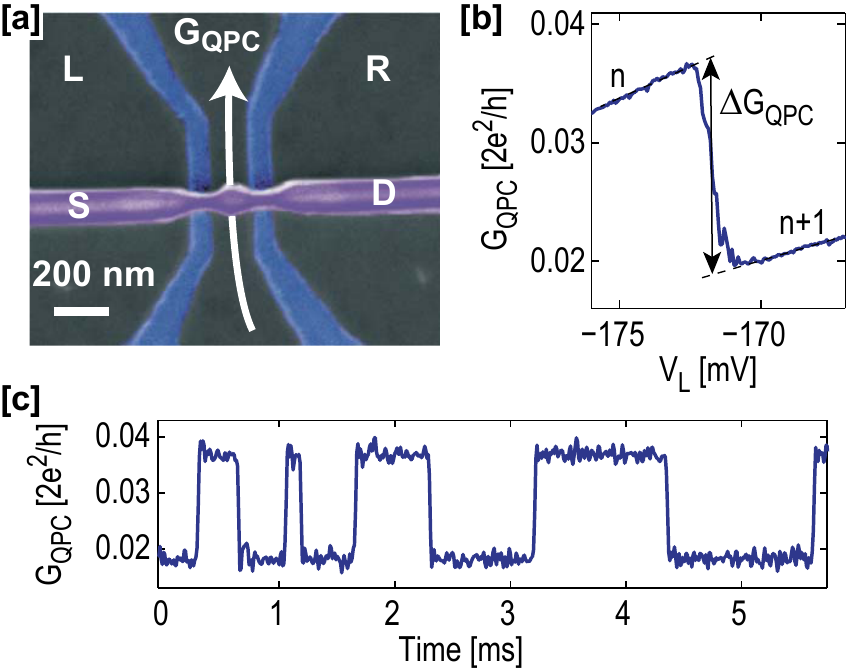}
 \caption{(color online) (a) SEM image of the device. The quantum dot is formed
 in the nanowire, with the quantum point contact located in the 2DEG
 directly beneath the QD. (b) QPC conductance measured versus voltage on gate L.
 At $V_\mathrm{L} = -172\mV$ an electron is added to the QD, leading to a decrease of $\Gqpc$.
 (c) Time trace of the QPC
 conductance measured at $V_\mathrm{L} = -172\mV$, showing a few electrons tunneling into and out of the
 QD. The upper level corresponds to a situation with $n$ electrons
 on the QD.
 }
\label{fig:sample}
\end{figure}

Figure~\ref{fig:sample}(a) shows a scanning electron microscope
(SEM) image of the device used in these experiments. An InAs
nanowire is deposited on top of a shallow (37 nm) AlGaAs/GaAs
heterostructure based two-dimensional electron gas (2DEG). The QPC
is defined by etched trenches, which separate the QPC from the rest
of the 2DEG. The parts of the 2DEG marked by L and R are used as
in-plane gates. The horizontal object in the figure is the nanowire
lying on top of the surface, electrically isolated from the QPC. The
QD in the nanowire and the QPC in the underlying 2DEG are defined in
a single etching step using patterned electron beam resist as an
etch mask. The technique ensures perfect alignment between the two
devices. Details of the fabrication procedure can be found in
Ref.~\onlinecite{shorubalko:2007}. The electron population of the QD
is tuned by applying voltages to the gates L and R. When changing
gate voltages, we keep the QPC potential fixed by applying a
compensation voltage $\Vtdeg$ to the 2DEG connected to both sides of
the QPC. All measurements presented here were performed at an
electron temperature of $T=2~\mathrm{K}$.

\section{Charge detection with a quantum point contact} \label{sec:QP_chargeDetection}
Figure~\ref{fig:sample}(b) shows a measurement of the QPC
conductance as a function of voltage on gate L. The gate voltage
tunes both the QPC transmission as well as the electron population
on the QD. The measurement was performed without any bias voltage
applied to the QD and with the drain lead of the QD pinched off. At
$V_\mathrm{L} = -172\mV$, the electrochemical potential of the QD
shifts below the Fermi levels of the source lead and an electron may
tunnel onto the QD. This gives a decrease $\Delta \Gqpc$ of the QPC
conductance corresponding to the change $\Delta q = \mathrm{e}$ of
the charge population on the QD. The curve in
Fig.~\ref{fig:sample}(b) shows the average QPC conductance giving
the time-averaged QD population. By performing a time-resolved
measurement, electron tunneling can be detected in real-time. This
is visualized in Fig.~\ref{fig:sample}(c), where the measured QPC
conductance fluctuates between the two levels corresponding to $n$
and $n+1$ electrons on the QD.  Transitions between the levels occur
on a millisecond timescale, which provides a direct measurement of
the tunnel coupling between the QD and the source lead
\cite{schleser:2004}. By analyzing the time intervals between
transitions, the rates for electrons tunneling into or out of the QD
can be determined separately \cite{gustavsson:2006}.

Next, we investigate the best regime for operating the QPC as a
charge detector.
The conductance of a QPC depends strongly on the confinement
potential $U_\mathrm{QPC}(\vec{r})$. When operating the QPC in the
region between pinch-off and the first plateau $(0 < G < 2e^2/h)$, a
small perturbation $\delta U_\mathrm{QPC}(\vec{r})$ leads to a large
change in conductance $\delta G$. If a QD is placed in close
vicinity to the QPC, we expect a fluctuation $\delta q$ in the QD
charge population to shift the QPC potential
$U_\mathrm{QPC}(\vec{r})$ and thus give rise to a measurable change
in QPC conductance. A figure of merit for using the QPC as a charge
detector is then
\begin{equation}\label{eq:QtoG}
\frac{\delta G}{\delta q}=  \frac{\delta
G\left[U_\mathrm{QPC}(\vec{r})\right]}{\delta
U_\mathrm{QPC}(\vec{r})} \, \frac{\delta
U_\mathrm{QPC}(\vec{r})}{\delta q}.
\end{equation}
The first factor describes how the conductance changes with
confinement potential, which depends strongly on the operating point
of the QPC. The second factor describes the electrostatic coupling
between the QD and the QPC and is essentially a geometric property
of the system.



\begin{figure}[tb]
\centering
\includegraphics[width=\linewidth]{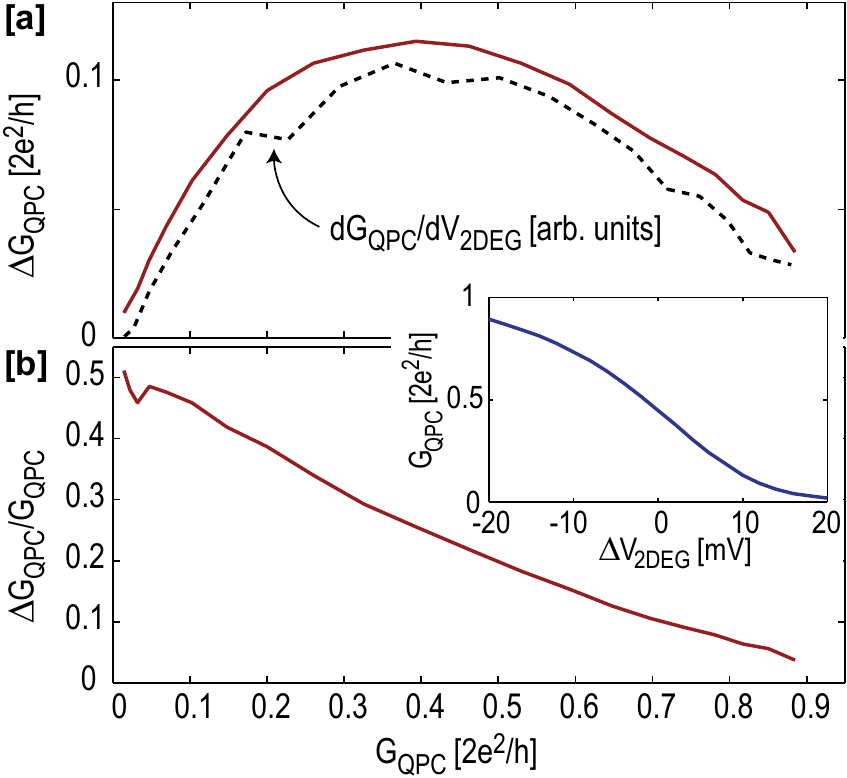}
\caption{(color online) (a) Change of QPC conductance as one
electron enters the QD, measured for different values of average QPC
conductance.  The dashed line is the numerical derivative of the QPC
conductance with respect to gate voltage.  The change is maximal at
$G_\mathrm{QPC} = 0.4 \times 2 e^2/h$, which coincides with the
steepest part of the QPC conductance curve [see inset in (b)].
 (b) Relative change of QPC conductance for one electron entering the QD, defined as
$(G_\mathrm{high}-G_\mathrm{low})/G_\mathrm{high}$. The relative
change increases with decreased $G_\mathrm{QPC}$, reaching above
50\% at $G_\mathrm{QPC} = 0.02 \times 2 e^2/h$. The inset shows the
variation of $G_\mathrm{QPC}$ as a function of gate voltage. }
\label{fig:StepVsG}
\end{figure}

The performance of the charge detector depends strongly on the
operating point of the QPC. The best sensitivity for a device of
given geometry is expected when the QPC is tuned to the steepest
part of the conductance curve. This corresponds to maximizing the
factor $\delta G / \delta U_\mathrm{QPC}$ in \EqRef{eq:QtoG}. In
\FigRef{fig:StepVsG}(a) we plot the conductance change $\Delta G$
for one electron entering the QD versus QPC conductance, in the
range between pinch-off and the first conductance plateau ($0 <
G_\mathrm{QPC} <2e^2/h$). The change $\Delta G$ is maximal around $
G_\mathrm{QPC} \sim 0.4 \times 2 e^2/h$ but stays fairly constant
over a range from 0.3 to $0.6 \times 2 e^2/h$. The dashed line in
\FigRef{fig:StepVsG}(a) shows the numerical derivative of $\Gqpc$
with respect to gate voltage. The maximal value of $\Delta G$
coincides well with the steepest part of the QPC conductance curve.
The inset in the figure shows how
the conductance changes as a function of gate voltage. 

In \FigRef{fig:StepVsG}(b), we plot the relative change in
conductance $\Delta G/G_\mathrm{QPC}$ for the same set of data. The
relative change increases monotonically with decreasing conductance,
reaching above 50\% at $G_\mathrm{QPC} = 0.02 \times 2 e^2/h$.
The relative change in QPC conductance $\Delta \Gqpc/\Gqpc$ in this
particular device is extraordinarily large compared to top-gate
defined structures, where $\Delta \Gqpc/\Gqpc$ is typically around
one percent for the addition of one electron on the QD
\cite{vandersypen:2004, reilly:2007}. We attribute the large
sensitivity to the close distance between the QD and QPC ($\sim \!
50\nm$, due to the vertical arrangement of the QD and QPC) and to
the absence of metallic gates on the heterostructure surface, which
reduces screening.

The results of Fig.~\ref{fig:StepVsG} indicate that it may be
preferable to operate the charge detector close to pinch-off, where
the relative change in conductance is maximized.
The quantity relevant for optimal detector performance in the
experiment is the signal-to-noise (S/N) ratio between the change in
conductance $\Delta G$ and the noise level of the QPC conductance
measurement. We measure the conductance by applying a fixed bias
voltage $\Vsd$ across the QPC and monitoring the current. In the
linear response regime, both the average current $\Iqpc$ and the
change in current for one electron on the QD ($\Delta \Iqpc$) scale
linearly with applied bias. The noise in the setup is dominated by
the voltage noise of the amplifier, which is essentially independent
of the QPC operating point and the applied bias in the region of
voltages discussed here. The S/N thus scale directly with $\Delta
\Iqpc$
\begin{equation}\label{eq:DeltaItoSN}
 S/N = \frac{\Delta \Iqpc^2}{\langle \Delta I^2_\mathrm{noise}\rangle}
 \propto \Vqpc^2  \, \Delta G^2  = \Iqpc^2  \left(\frac{\Delta
 G}{G}\right)^2.
\end{equation}
In practice the maximal usable QPC current is limited by effects
like heating or emission of radiation which can influence the
measured system. If we consider limiting the current, we see from
Eq.~(\ref{eq:DeltaItoSN}) that the highest S/N is reached for the
maximal value of $\Delta G/G$ at $\Gqpc \ll 2e^2/h$. However, this
operation point requires a large voltage bias to be applied to the
QPC. If the QPC bias is larger than the single-particle level
spacing of the QD, the current in the QPC may drive transitions in
the QD and thus exert a back-action on the measured device
\cite{gustavssonPRL:2007} (see next section). Therefore, a better
approach is to limit the QPC voltage. Here, the best S/N is obtained
when optimizing $\Delta G$ rather than $\Delta G/G$ and operating
the QPC close to $\Gqpc=0.5\times 2 e^2/h$. The sensitivity of the
QPC together with the bandwidth of the measurement circuit allows a
detection time of around $4~\mu s$ \cite{gustavssonAPL:2008}. The
tunneling rates presented in the following were extracted taking the
finite detector time into account \cite{naaman:2006}.

\section{Excitations driven by the quantum point contact}
In this section we study QD transitions driven by current
fluctuations in the QPC. Such excitations were already studied for
QDs and QPCs defined in a GaAs 2DEG
\cite{onacQD:2006,khrapai:2006,gustavssonPRL:2007}. From those
experiments, it was not clear how energy was mediated between the
systems. The nanowire sample investigated here is conceptionally
different because the QD and the QPC are fabricated in different
material systems. This allows us to make a statement about the
physical processes involved in transmitting energy between the QD
and the QPC. Since the two structures sit in separate crystals with
different lattice constants and given that the systems hardly touch
each other, we can assume that phonons only play a minor role as a
coupling mechanism.
Instead, we assume the QD transitions to be driven by radiation
emitted from the QPC \cite{aguado:2000}. Another advantage of the
nanowire structure compared to GaAs systems is that the QD energy
scales are an order of magnitude larger compared to QDs formed in a
GaAs 2DEG. This allows us to investigate radiation at several
$100\GHz$, reaching into the long-wavelength infrared regime.

\begin{figure}[tb] \centering
\includegraphics[width=\linewidth]{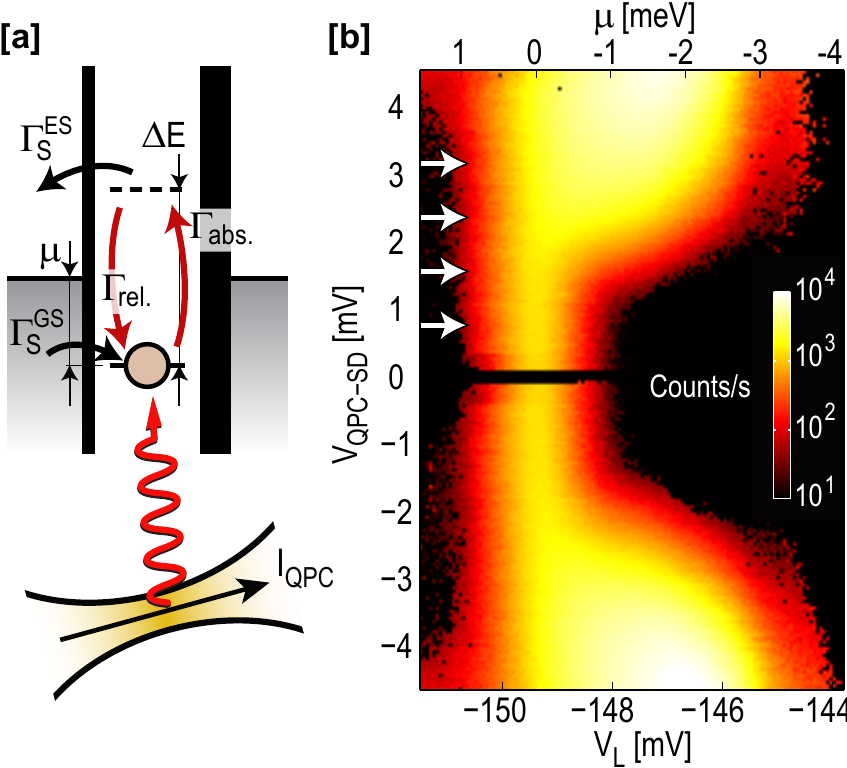}
\caption{ (a) Energy level diagram describing the absorption
process.
The electron in the QD is excited due to photon absorption, which allows it to tunnel out to the lead. 
(b) Quantum dot excitations, measured versus QPC bias voltage. The
main peak at $\mu=0$ is due to equilibrium fluctuations between the
source lead and the QD. As the gate voltage $V_\mathrm{L}$ is
increased, the electrochemical potential of the QD drops below the
Fermi level of the lead and only tunneling processes involving QD
excitations become possible. Photon absorption is only possible for
QPC bias voltages higher than the QD level separation $\Delta E$,
giving rise to the shoulder-like features appearing at high $\Vqpc$.
The data was extracted from QPC conductances traces taken at
$\Gqpc=0.4\times 2e^2/h$, filtered at $50\kHz$. The data taken at
low QPC bias $|\Vqpc|<0.4\mV$ was filtered at a lower bandwidth
($15\kHz$) to allow counting in this regime.}
 \label{fig:ShoulderMap}
\end{figure}

We first discuss the QD configuration used for probing the radiation
of the QPC. Since the QD level spectrum is not tunable, we can only
drive transitions at fixed frequencies corresponding to excited
states in the QD \cite{onacQD:2006}. Figure~\ref{fig:ShoulderMap}(a)
shows the level configuration of the system, with the QD
electrochemical potential $\mu$ below the Fermi level of the leads.
The tunneling barriers are highly asymmetric, with the barrier
connecting the QD to the drain lead being almost completely pinched
off. We do not apply any bias voltage to the QD. The system is in
Coulomb blockade, but by absorbing a photon the QD may be put into
an excited state with electrochemical potential above the Fermi
energy of the leads. From here, the electron may leave to the source
contact, the QD is refilled and the cycle may be repeated.

In Fig.~\ref{fig:ShoulderMap}(b) we plot the electron count rate
versus QD potential and QPC bias. The peak at $\mu=0$ is due to
equilibrium fluctuations between the QD and the source contact, with
the width set by the electron temperature in the lead. As the QPC
bias is increased above $\approx 2.5\mV$, a shoulder appears in the
region of $\mu<0$. This is consistent with the picture in
Fig.~\ref{fig:ShoulderMap}(a); we need to apply a QPC bias larger
than the single-level spacing for the photon-assisted tunneling to
become possible. The width of the shoulder is set by $\Delta E
\approx 2.5\meV$ and is therefore expected to be independent of QPC
bias; we will see later in this section that the apparent smearing
of the features in Fig.~\ref{fig:ShoulderMap}(b) are due to
temperature and tuning of the tunneling rates. The picture is
symmetric with respect to $\Vqpc$, meaning that the emission and
absorption processes do not depend on the direction of the QPC
current. The lack of data points around $\Vqpc=0$ are due to the
fact that the low QPC bias prevents the operation of the QPC as a
charge detector. Due to the asymmetric coupling of the QD to the
source and drain lead, we could not make a direct confirmation of
the existence of an excited state with $\Delta E = 2.5\mV$ using
finite bias spectroscopy. However, the value is consistent with
excited states found in Coulomb diamond measurements in regimes
where the tunnel barriers are more symmetric
\cite{gustavssonAPL:2008}.

\begin{figure}[tb] \centering
\includegraphics[width=\linewidth]{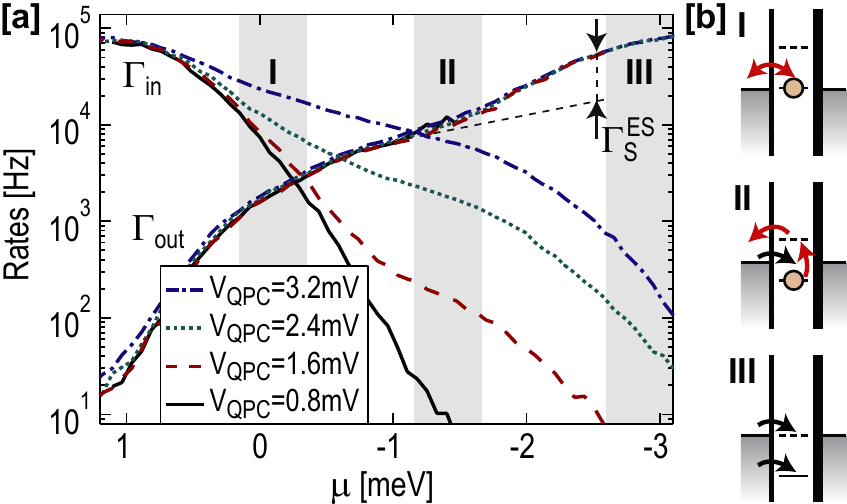}
\caption{ (a) Rates for electrons tunneling into and out of the QD,
measured at four cross-sections of Fig.~\ref{fig:ShoulderMap}(b)
[position of arrows in Fig.~\ref{fig:ShoulderMap}(b)]. Only the rate
related to absorption ($\Gout$) is strongly influenced by the
increase in QPC bias. (b) Energy level diagrams for the three
configurations marked in (a).}
 \label{fig:ShoulderTrace}
\end{figure}

Figure \ref{fig:ShoulderTrace}(a) shows the separate rates for
electrons tunneling into and out of the QD at horizontal
cross-sections of Fig.~\ref{fig:ShoulderMap}(b), measured at four
different QPC bias voltages [marked by arrows in
Fig.~\ref{fig:ShoulderMap}(b)]. Around the resonance [$\mu=0$, case
I in Fig.~\ref{fig:ShoulderTrace}], the tunneling is due to
equilibrium fluctuations and the rates for tunneling into and out of
the QD are roughly equal. By lowering the electrochemical potential
$\mu$ the rate for electrons leaving the QD first falls off
exponentially due to the thermal distribution of the electrons in
the lead. Continuing to case II of Fig.~\ref{fig:ShoulderTrace}, we
come into the regime of QD excitations. Here, the rate $\Gout$ is
directly related to the absorption process sketched in
Fig.~\ref{fig:ShoulderMap}(a), while the rate $\Gin$ corresponds to
the refilling of an electron from the lead. Consequently, $\Gout$
shows a strong QPC bias dependence, while $\Gin$ stays roughly
constant.

In case III, the excited state goes below the Fermi level of the
source lead and the absorption rate drops quickly. At the same time,
$\Gin$ increases as the refilling of an electron into the QD may
occur through either the ground state or the excited state. This
provides a way to determine the tunnel coupling between the source
contact and the excited state in the QD ($\Gs^\mathrm{ES}$). From
the data in Fig.~\ref{fig:ShoulderTrace}(a), we estimate
$\Gs^\mathrm{ES} \approx 60\kHz-20\kHz = 40\kHz$. The change of
tunnel coupling with gate voltage makes the exact determination of
$\Gs^\mathrm{ES}$ difficult, the value given here should only be
considered as a rough estimate.

The tunneling rates within the region of photon-assisted tunneling
are strongly depending on gate voltage. Similar effects have been
investigated in 2DEG QDs, where the tunneling rate of a barrier was
shown to depend exponentially on gate voltage due to tuning of the
effective barrier height \cite{maclean:2007}. A difference of our
sample compared to 2DEG QDs concerns the properties of the
electronic states in the leads. For GaAs QDs, the leads consist of a
two-dimensional electron gas where the ideal density of states (DOS)
is independent of energy. For the nanowire QD, the leads are also
parts of the nanowire and the corresponding electron DOS may show
strong variations with energy due to the quasi-one dimensionality
and finite length of the wire. Within the region of photon-assisted
tunneling in Fig.~\ref{fig:ShoulderTrace}(a), we shift the
electrochemical potential of the QD and thereby change the energy of
the tunneling electrons relative to the Fermi level in the lead. The
measured tunneling rates could therefore show variations due to
changes in the DOS in the lead.

However, the behavior seen in region II of
Fig.~\ref{fig:ShoulderTrace}(a) is not compatible with the effects
discussed in the previous paragraph. The rate $\Gin$ is directly
related to the tunnel coupling $\Gs^\mathrm{GS}$ between the source
lead and the QD ground state, while the rate $\Gout$ depends on the
coupling $\Gs^\mathrm{ES}$ between source and the excited state in
the QD. For arguments based on barrier tuning and varying electron
DOS, we would expect both $\Gs^\mathrm{GS}$ and $\Gs^\mathrm{ES}$ to
change in the same way with gate voltage. This is in disagreement
with the results of Fig.~\ref{fig:ShoulderTrace}(a); $\Gin$
\emph{increases} while $\Gout$ \emph{decreases} with gate voltage.
Instead, we speculate that the observed behavior may be due to
non-resonant processes involving energy relaxation in the leads.
Focusing on the energy level configuration pictured in
Fig.~\ref{fig:ShoulderMap}(a), we see that there are a large number
of occupied states in the lead with energy higher than the
electrochemical potential of the QD. Elastic tunneling into the QD
can only occur for electrons with energy equal to the
electrochemical potential of the QD, but electrons at higher energy
may contribute to the measured rate in terms of processes involving
relaxation. As we lower $\mu$, the number of initial states
available for the inelastic processes increase and would therefore
explain the \emph{increase} in $\Gin$ with decreased $\mu$.
Inelastic tunneling is also possible for electrons leaving the QD
excited state to empty states in the lead. Here, the number of empty
states available for the inelastic processes goes down when the QD
potential is lowered. This is in agreement with the measured
\emph{decrease} in $\Gout$ with decreased $\mu$.

\section{QPC bias dependence}\label{sec:NW_PATvsBias}
Next, we investigate how the QPC bias influences the efficiency of
the photon absorption process. For this purpose we apply a
rate-equation model similar to that used for investigating
QPC-driven excitations in double QDs \cite{gustavssonPRL:2007}. The
model consists of three states, corresponding to the QD being empty,
populated with an electron in the ground state, or populated with an
electron in the excited state. We write down the master equation for
the occupation probabilities $p =
[p_\mathrm{GS},~p_\mathrm{ES},~p_0]$ of the three states
\begin{equation}\label{eq:master} \frac{d}{dt }\left[\!
\begin{array}{c}
             p_\mathrm{GS} \\
             p_\mathrm{ES} \\
             p_0 \\
              \end{array} \!\right] =
                        \left[\!
          \begin{array}{ccc}
            -\Gabs & \Grel & \Gs^\mathrm{GS} \\
            \Gabs & -(\Gs^\mathrm{ES}+\Grel) & 0 \\
            0 & \Gs^\mathrm{ES} & -\Gs^\mathrm{GS} \\
          \end{array}
        \!\right] \left[\!
                        \begin{array}{c}
             p_\mathrm{GS} \\
             p_\mathrm{ES} \\
             p_0 \\
                        \end{array}
                      \!\right].
\end{equation}
Here, $\Gabs$ is the absorption rate and $\Grel$ is the relaxation
rate of the QD. The rates are visualized in
\FigRef{fig:ShoulderMap}(a).
The charge detection technique can only probe rates for electrons
entering or leaving the QD. These rates are found from the
steady-state solution of \EqRef{eq:master}:
\begin{eqnarray}
  \nonumber \Gin &=& \Gs^\mathrm{GS}, \\
  \Gout &=&  \Gs^\mathrm{ES} \, \frac{p_\mathrm{ES}}{p_\mathrm{ES}+p_\mathrm{GS}} =
  \Gs^\mathrm{ES} \, \frac{\Gabs}
  {\Gs^\mathrm{ES} + \Gabs + \Grel}. \label{eq:GoutEq}
\end{eqnarray}
In GaAs QDs, the charge relaxation process occurs on a timescale of
$\sim\!10~\mathrm{ns}$ \cite{fujisawa:2002}. Similar rates are
expected for nanowire QDs. Therefore, we assume $\Gs^\mathrm{ES} \ll
\Grel$ and estimate the behavior of $\Gout$ in the limit of weak
absorption ($\Gabs \ll \Grel$). Here, \EqRef{eq:GoutEq} simplifies
to
\begin{equation}\label{eq:GoutEqGrel}
  \Gout  = \Gs^\mathrm{ES} \, \Gabs/\Grel.
\end{equation}
Under these conditions the measured rate $\Gout$ is expected to
scale linearly with the absorption rate. Assuming the excitations to
be driven by fluctuations in the QPC current, we can combine
\EqRef{eq:GoutEqGrel} with the QPC emission spectrum $S_I(\omega)$
\cite{aguado:2000},
\begin{eqnarray}\label{eq:GoutSI}
  \Gout  &\propto& \Gs^\mathrm{ES} \, S_I(\Delta E/\hbar)
   =  \nonumber \\
   &=& \Gs^\mathrm{ES} \,
\frac{4 e^2}{h} D (1-D) \frac{e V_\mathrm{QPC} - \Delta E}{1-e^{-(e
V_\mathrm{QPC} -  \Delta E)/k_B T}},
\end{eqnarray}
where $D$ is the QPC transmission coefficient and $T$ the electron
temperature in the QPC leads. Note that \EqRef{eq:GoutSI} only gives
the proportionality between $\Gout$ and $S_I(\omega)$; to make
quantitative predictions for the absorption rate one needs to
determine the overlap between the ground and the excited state. For
a double QD, this coupling can be extracted from charge localization
measurements \cite{dicarlo:2004}. However, it is not as
straightforward to estimate the overlap for single-QD excitations.
One would need to know the shape of the wavefunctions for the
different QD states, which is not known.

\begin{figure}[tb] \centering
\includegraphics[width=\linewidth]{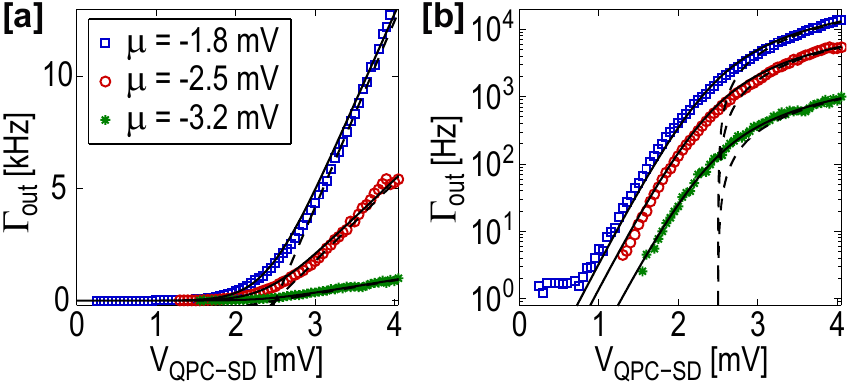}
\caption{(a) Electron tunneling due to photon absorption, measured
versus QPC bias voltage. The data was taken at three different
positions of the shoulder seen in Fig.~\ref{fig:ShoulderMap}. As
soon as the QPC bias voltage exceeds the QD level separation $\Delta
E$, the absorption rate increases linearly with $\Vqpc$. (b) Same
data as in (a), but plotted in logarithmic scale. The absorption
rate shows exponential decay for $e\Vqpc<\Delta E$, with the slope
of the decay set by the electron temperature in the QPC. The solid
lines are fits to \EqRef{eq:GoutSI} in the text, while the dashed
lines are the corresponding results assuming zero temperature. }
 \label{fig:ShoulderBias}
\end{figure}

In Fig.~\ref{fig:ShoulderBias} we plot the measured tunneling rate
$\Gout$ related to absorption versus bias on the QPC, measured for
three different electrochemical potentials of the QD. The traces
correspond to vertical cross-sections for positive $\Vqpc$ in
Fig.~\ref{fig:ShoulderMap}(b). Figure~\ref{fig:ShoulderBias}(a)
shows the rates plotted on a linear scale; the rates taken at all
three positions increase linearly with QPC bias as soon as
$e\Vqpc>\Delta E$. The solid lines are fits to \EqRef{eq:GoutSI}
with $T=2\K$ and assuming $\Delta E = 2.5\meV$ to be the same for
all three traces. As described in the previous section, we attribute
the difference in slope for the three cases to changes in effective
tunnel coupling with gate voltage [see
Fig.~\ref{fig:ShoulderTrace}(a)]. Figure~\ref{fig:ShoulderBias}(b)
shows the same data plotted on a logarithmic scale. Here, we see a
clear exponential decay for $e\Vqpc<\Delta E$; this is due to the
thermal distribution of electrons in the QPC \cite{zakka:2007}. The
dashed lines in Fig.~\ref{fig:ShoulderBias} show the rates expected
for the case of zero temperature. The weak but non-zero count rate
occurring at low QPC bias voltages ($\Vqpc<1\mV$) for the data taken
at $\mu=-1.8\ueV$ is due to non-photon induced thermal fluctuations
between the QD ground state and the lead.

To quantify the efficiency of the absorption process we compare the
rates $\Gout$ and $\Gs^\mathrm{ES}$ using \EqRef{eq:GoutEqGrel}. Due
to the strong change of $\Gs^\mathrm{ES}$ with gate voltage, we can
only make a quantitative comparison between $\Gout$ and
$\Gs^\mathrm{ES}$ in the region where we are able to determine
$\Gs^\mathrm{ES}$ (around $-\mu=\Delta E$). This corresponds to the
circles in Fig.~\ref{fig:ShoulderBias}. For this data set, the
measured rate goes up to around $5\kHz$ for $\Vqpc=4\mV$, so that we
still have $\Gout \ll \Gs^\mathrm{ES}$.
This confirms that we are in a regime of weak absorption where the
relative population of the excited state is much smaller than the
population of the ground state. Note that the same is most likely
true also for the data sets taken at $\mu=-1.8\mV$ and $\mu=-3.2\mV$
in Fig.~\ref{fig:ShoulderBias}; however, we can not make a
quantitative comparison with $\Gs^\mathrm{ES}$ since we do not have
an independent measurement of $\Gs^\mathrm{ES}$ for those regions.


\section{Changing the QPC operating point}\label{sec:PATvsG}
In this section, we modify the operating point of the QPC to check
how this influences properties of the emitted radiation. Since we
use the same QPC both for emitting radiation and for performing
charge detection, it is not possible to operate the device at the
plateaus where the conductance is fully quantized. However, we could
tune the QPC conductance in a region between $0.05\times 2e^2/h <
\Gqpc < 0.8 \times 2e^2/h$ while still being able to detect the
tunneling electrons.

\begin{figure}[tb] \centering
\includegraphics[width=\linewidth]{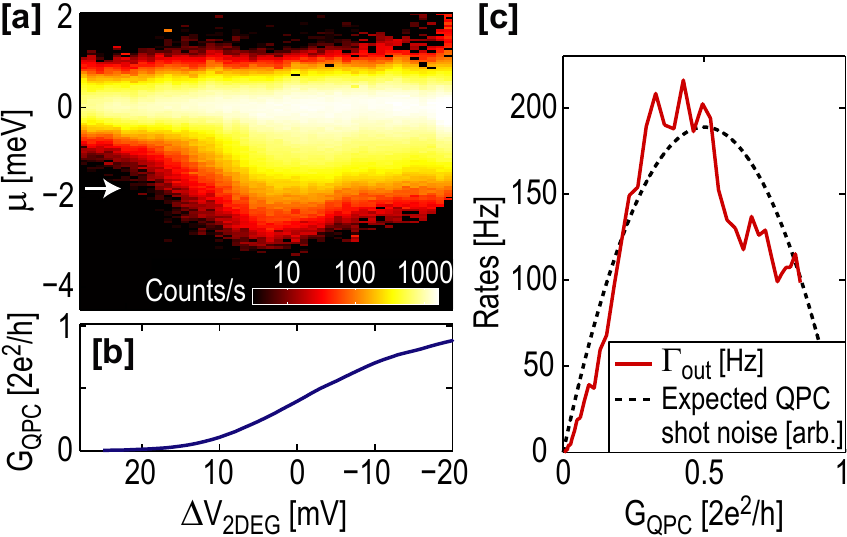}
\caption{
 (a) Electron count rate in the regime of the absorption process,
measured versus gate voltage on the 2DEG. This corresponds to tuning
the conductance of the QPC. A vertical cross-section corresponds to
the shoulder seen in Fig.~\ref{fig:ShoulderMap}. The data was taken
with $\Vqpc = 2\mV$.
 (b) QPC conductance as a function of gate voltage, measured for the
same region as in (a).
 (c) Tunneling rate $\Gout$ associated with the absorption process,
measured at $\mu=-1.8\meV$ [marked by an arrow in (a)]. The dashed
line shows the expected emission spectrum of the QPC, up to a
scaling factor depending on the efficiency of the absorption
process.}
 \label{fig:ShoulderGqpc}
\end{figure}

In Fig.~\ref{fig:ShoulderGqpc}(a) we plot the electron count rate at
the photon-absorption shoulder versus change in $\Vtdeg$. Figure
\ref{fig:ShoulderGqpc}(b) shows how the QPC conductance changes with
gate voltage within the region of interest. Compensation voltages
were applied to the gates L and R in order to keep the QD potential
fixed while sweeping $\Vtdeg$. The data was taken with fixed $\Vqpc
= 2\mV$ to make the photon absorption process possible. This bias is
still lower than the characteristic sub-band spacing of the QPC,
which is around $5\meV$.
The strong peak at the top of Fig.~\ref{fig:ShoulderGqpc}(a)
($\mu=0$) corresponds to equilibrium fluctuations between the QD and
the source lead. In the region of photon-assisted tunneling [marked
by the arrow in Fig.~\ref{fig:ShoulderGqpc}(a)], the shoulder
appears with increasing QPC conductance. Going above $\Gqpc = 0.5
\times 2e^2/h$, the strength of tunneling at the position of the
shoulder decays slightly.

Assuming that the shoulder appears because of radiation emitted from
shot noise fluctuations in the QPC current, we expect the measured
absorption rate to depend on the transmission of the QPC. From
\EqRef{eq:GoutSI} we see that the emission spectrum scales with
$D(1-D)$, where $D$ is the transmission coefficient of the channel.
In Fig.~\ref{fig:ShoulderGqpc}(c) we plot the rate $\Gout$ related
to the absorption process, measured at $\mu=-1.9\meV$ [position of
the arrow Fig.~\ref{fig:ShoulderGqpc}(a)]. The dashed line shows the
emission expected from the QPC, $S_I \propto D(1-D)$. For low
$\Gqpc$, the measured rate follows the expected emission spectrum
reasonably well, with a maximum around $\Gqpc \approx 0.5 \times
2e^2/h$.

Still, the measured curve shows deviations compared to the predicted
behavior. Suppression of noise close to $G_\mathrm{QPC} = 0.7 \times
2 e^2/h$ has been reported \cite{roche:2004,dicarlo:2006} to be
related to 0.7 anomaly \cite{thomas:1996}. There are indications of
the 0.7 anomaly also in our sample, but we believe the deviations in
the measured noise spectrum are more likely to originate from an
increase in background charge fluctuations triggered by the QPC
current. At large QPC currents ($\Iqpc>20\nA$), the noise in the
system increases with $\Iqpc$. This can not be attributed to the
intrinsic QPC shot noise but is rather due to fluctuations of
trapped charges driven by the high QPC current. The QD is thus
placed in an environment of fluctuating potentials, which may lead
to QD transitions. The strength of such transitions depends strongly
on the number of fluctuators in the neighborhood of the QD
\cite{pioda:2007}. The charge traps also influence the count rate in
the regime of tunneling due to equilibrium fluctuations [peak at
$\mu=0$ in Fig.~\ref{fig:ShoulderGqpc}(a)]. For $\Gqpc=0.8 \times
2e^2/h$ ($\Delta \Vtdeg=-20\mV$), the peak is considerably wider
than for $\Gqpc=0.05 \times 2e^2/h$. Again, this can be attributed
to a fluctuating potential at the location of the QD.

To minimize the influence of the charge traps, one would prefer to
decrease $\Vqpc$ and operate the QPC at lower current levels.  For
the configuration used in Fig.~\ref{fig:ShoulderGqpc}
($\Vqpc=2\mV$), the QPC current reaches values above $100\nA$ at
$\Gqpc \approx 0.7 \times 2e^2/h$. However, $\Vqpc$ can not be made
too small; we need to make sure that $e\Vqpc$ is on the same order
of magnitude as the level spacing $\Delta E$, otherwise the QPC will
not emit radiation in the right frequency range. The above
discussion only concerns a measurement of the emission properties of
the QPC. When using the device to probe radiation of an external
source, the QPC can be operated at much lower bias voltages.


To summarize, we have used time-resolved charge detection techniques
to investigate the influence of current flow in a near-by QPC to the
electron population in nanowire QD. Since the QD and the QPC are
fabricated in different material systems, we conclude that phonons
can only play a minor role for a transferring energy between the
structures. Instead, we attribute the charge to absorption of
photons emitted from the quantum point contact. The large energy
scales of the nanowire QD allows detection of radiation at a
frequency of $f = 2.5\meV/h = 0.6~\mathrm{THz}$, thus reaching into
the long-wavelength infrared regime. This is an order of magnitude
larger than energy scales reachable with GaAs QDs
\cite{gustavssonPRL:2007}.

\bibliographystyle{apsrev}
\bibliography{QPCNanowire}

\end{document}